\documentstyle[12pt,a4]{article}
\begin{document}
\titlepage
\baselineskip 27pt
\begin{flushright}
IMSc -- 96/02/04   \\
February, 1996
\end{flushright}
\bigskip
\vspace{1in}
\begin{center}
{\bf Field Theory of Quantum Antiferromagnets: }\\
{\bf From The Triangular To The Kagome Lattice}
\end{center}
\vspace{0.9in}
\begin{center}
D.Shubashree \footnote{email: shuba@imsc.ernet.in} 
and R.Shankar\footnote{email:shankar@imsc.ernet.in} \\

The Institute of Mathematical Sciences, \\
C. I. T. Campus, Taramani\\
Madras 600113
\end{center}
\bigskip
%\newpage
%\setcounter{page}{0}
\thispagestyle{empty}
\vspace{2in}
\begin{center}
{\bf ABSTRACT}\\
\end{center}

We analyse a family of models,
that interpolates between the Triangular lattice
antiferromagnet (TLAF) and the Kagome lattice
antiferromagnet (KLAF) . 
 We identify the field theories governing the low energy, long 
wavelength physics of these models .
Near the TLAF the low
energy field theory is a nonlinear sigma model of a $SO(3)$ group valued
field. The  $SO(3)$ symmetry of the spin system  is enhanced to a 
$SO(3)_R \times SO(2)_L$ symmetry in the field theory. Near the KLAF other
modes become important and the field takes values in $SO(3) 
\times S_2$. We analyse this field theory and show that it  
admits a novel phase in which the $SO(3)_R$ spin symmetry is 
unbroken and the $SO(2)_L$ symmetry is broken. 
We propose this as a possible mechanism by which a gapless excitation 
can exist in the KLAF without breaking the spin rotation symmetry .

\newpage
%%%list of newcommands%%%%%%%%%%%%%%%%%%%%%%%%%%%
 \newcommand{\nc}{\newcommand}
\nc{\expw}{\mbox{$ \exp{- \frac{i w_{Ii\alpha}}{\sqrt{\tilde s}}}$}}
\nc{\expwd}{\mbox{$ \exp{ \frac{i w_{Ii\alpha}}{\sqrt{\tilde s}}}$}}
\nc{\expn}{\mbox{$ \exp{- i m_{I}\phi r }$}}
\nc{\expwk}{\mbox{$ \exp{- \frac{i w_{k,i,\alpha}}{\sqrt{\tilde s}}}$}}
%\nc{\expwkk}{\mbox{$ \exp{- \frac{i w_{-k,i,\alpha}}{\sqrt{\tilde s}}}$}}
\nc{\ki}{\mbox{$\chi$}}
\nc{\nal}{n_{\alpha}}
\nc{\nbet}{n_{\beta}}
\nc{\beq}{\begin{equation}} 
\nc{\eeq}{\end{equation}}
\nc{\beqar}{\begin{eqnarray}} 
\nc{\eeqar}{\end{eqnarray}}
\nc{\wIi}{\mbox{$w_{Ii\alpha}$}}
 \nc{\wJj}{\mbox{$w_{Jj\beta}$}}
\nc{\wki}{\mbox{$w_{Ki\alpha}$}} 
\nc{\wkki}{\mbox{$w_{-Ki\alpha}$}}
\nc{\wkkj}{\mbox{$w_{-Kj\beta}$}}
 \nc{\cn}{\mbox{$c_{n r }$}}
\nc{\cpn}{\mbox{$c^{'}_{n r }$}} \def\src{SrCr$_8$Ga$_4$O$_{19}$}
%%%%%%%%%%%%%%%%%%%%%%%%%%%%%%%%%%%%%%%%%%%%%%%%%%%%%%%%%%%%%%%%%%%%
\section{Introduction.} The possibility of novel groundstates has been
motivating the study of two dimensional frustrated quantum
antiferromagnets for quite some time now. In classical unfrustrated
antiferromagnets, the ground state is the well known Neel state. The SO(3)
spin symmetry of the system is broken down to SO(2). The low energy
excitations are the two branches of gapless spin waves which are the
Goldstone modes. A commonly observed effect of frustration is that the
ground state becomes a spiral state.  The spin arrangement remains
periodic and is characterized by a spiral vector, {\bf q}.  The SO(3)
symmetry is now completely broken and there are three gapless Goldstone
modes. If the parameters of the system are such that the effects of
quantum fluctuations are not very large then this basic picture remains
true in the quantum system with changed values of physical quantities like
staggered magnetization, spin wave velocities etc.  However strong quantum
fluctuations can destroy the long range spin order and the system could go
to a paramagnetic phase. In frustrated systems several alternate novel
effects of quantum fluctuations have been proposed. One possibility  
\cite{pwa} is the spin liquid groundstate that has the full symmetry of the
hamiltonian. Closely related states are the chiral spin liquids or flux 
phases\cite{frad}. More recently,  magnetic states characterized by long range 
order of higher tensor operators have been proposed \cite {cc}. 

Numerical and analytical studies \cite {huse,rrps} indicate that the
triangular lattice antiferromagnet (TLAF) has a Neel ordered spiral ground
state with a spiral angle of $2 \pi / 3$ . This is the so called
$\sqrt{3}\times \sqrt{3} $ state. However, work on the spin 1/2 Kagome
lattice antiferromagnet (KLAF) indicates the absence of any kind of long
range Neel order \cite {zeng,chalk,leung}. The KLAF is therefore a potential 
candidate for novel groundstates. 

The KLAF is experimentally realized in the magnetoplumbite type compound
Sr Cr$_{8}$ Ga$_{4}$ O$_{19} $ \cite {obr}. This is a layered compound
containing planes of Cr$^{3+}$ ions that form a $S={3 \over 2}$ KLAF.
About $80\%$ of the KLAF sites are occupied by the chromium ions. The
inter Cr spacing is $2.9~~ A$. Susceptibility measurements show a Curie
-Weiss behaviour at high temperature with a Curie-Weiss temperature
$\theta _ {CW} \sim 400 K$. There is a spin glass like cusp at $T_{g} \sim
5 K$. The specific heat shows a $T^{2}$ behaviour below $T_{g}$ \cite
{ram}. Neutron scattering however shows no Bragg peaks down to 1.5K \cite
{broh}. There exists short range $\sqrt 3 \times \sqrt 3$ order with a
correlation length of about $7 A$ at $1.5 K$. This has led to the
speculation that the groundstate is characterized by the long range order
of some order parameter that is invisible to the neutrons. Recent $\mu\mbox{sr}$
 studies on the compound lends support to a spin liquid type of ground
state \cite { musr}. 
 
Recently another system where the KLAF is experimentally realized has
been reported
\cite{jaro}. This is the deuteronium jarosite,
(D$_3$O)Fe$_3$(SO$_4$)(OD)$_6$. The Fe$^{3+}$ atoms
in this compound form layers of $S={5 \over 2}$ KLAF. About $97\%$ of
the KLAF sites are 
occupied
by the iron ions. The inter Fe spacing is $3.67~~ A$. The Curie-Weiss
temperature is $\sim
1500~ K$. There is a spin glass type cusp at $13.8~ K$. The specific
heat goes as $T^2$ below
this temperature. Neutron scattering sees no long range order. There
is short range order
corresponding to the $\sqrt 3 \times \sqrt 3$ spin structure with a
correlation length ofabout $19~ A$ at $1.9~ K$.
 
The $T^2$ behaviour of the low temperature specific heat, absence of long 
range spin order and the presence of short range ${\sqrt 3} \times {\sqrt
3}$ order are common properties of both these systems indicating that
these are universal properties of a KLAF. The specific heat behaviour
indicates the presence of a gapless boson in the low temperature phase.
However the neutron scattering shows absence of long range spiral order.
Further, as mentioned above, numerical work indicates that all the
symmetries of the hamiltonian are intact. What is the mechanism in these 
systems that produces a gapless boson while keeping the symmetries of the 
hamiltonian intact ? In this paper, we address this puzzle and propose a 
possible solution for it. We work within the framework of the  large $S$ 
semiclassical expansion. The fairly high value of the spin in the experimental
systems indicates that these properties should be seen in this approximation.

The classical KLAF has infinitely many
(apart from symmetry operations) degenerate groundstates
including many with non-coplanar spin configurations. It 
exhibits the order from disorder phenomenon, i.e , the spin wave
modes around the planar groundstates are softer,  hence  the 
fluctuations partially lift the groundstate degeneracy. However,
there still remain infinitely many distinct planar ground state
spin configurations \cite {shend,chub}.

This property of the KLAF  makes it 
difficult to study analytically . In this paper we consider instead,
a one parameter family of models that interpolate between
the TLAF and the KLAF. Such a model has also been considered by 
Zeng and Elser in \cite {zeng},
where they do a spin wave analysis of the model .  
We will refer to these models as the deformed 
triangular lattice antiferromagnet (DTLAF). The model is defined
by the Hamiltonian,

\beq H = J~\big ( \sum_{<i,j>\epsilon K_{B}} \vec{S_{i}}.\vec{S_{j}}
 + \chi \sum_{<i,j>\not \epsilon K_{B}} \vec{S_i}. \vec{S_{j}} ~~\big )
\eeq

Here $ <i,j>$ label the nearest neighbour sites on a triangular lattice.
$K_{B}$ denotes the set of nearest neighbour bonds that belong to the
kagome lattice (which is a subset of the triangular lattice ).
When $\chi =1$, the model is the TLAF, whereas
if $\chi = 0$, it is the KLAF. It is also interesting to note that
the structure of the Cr atoms in SCGO is made up of a two layers .The
atoms in one plane lie on a Kagome lattice, while those on the upper layer
lie on a triangular lattice whose lattice points lie over the centres of
the hexagons in the kagome structure \cite{obr}. Therefore the DTLAF could
be of direct relevance to SCGO. 

An important property of the DTLAF which we will show in the next section 
is that the ground state is unique (upto symmetry operations)
for all nonzero values of \ki . For $0< \chi \leq 2 $,
the ground state is the $\sqrt{3} \times \sqrt{3} $ state . 
Our strategy is then to study the model at $\chi \neq 0$ and 
analyse the quantum groundstate as a function of $\chi$.
As mentioned earlier, short range $\sqrt{3} \times \sqrt{3}$ order has been 
experimentally observed both in \src $~ \mbox{and}$ in  
(D$_3$O)Fe$_3$(SO$_4$)(OD)$_6$. Theoretically also in a large N formalism,
the fluctuations pick out the $\sqrt{3} \times \sqrt{3}$ state \cite {
sach}. This indicates that it should be meaningful to look upon
the KLAF as the $\chi \rightarrow 0$ limit of the DTLAF.

The analysis of a spin system near a transition requires consideration of
large amplitude fluctuations. Further since the
correlation length near the transition is large, the lattice model
can be approximated by a continuum field theory . Thus the physics is
expected to be well described by a field theory of the softmodes of the
system . This expectation has been well verified experimentally for the
unfrustrated square lattice antiferromagnet where the physics is described
well by the nonlinear sigma model \cite {nel} . The order parameter for
this model is a unit vector field . The soft modes are the two Goldstone
modes. 

Field theories for frustrated systems, in particular for the TLAF,
have been derived \cite {domb} and analysed using momentum space
renormalization group techniques \cite {fried,az1,az2}. The order 
parameter here is a SO(3) group element . Physically a rotation group 
element can be looked upon as describing the orientation of a rigid body. 
In the spin system this orientation is specified by the sublattice
magnetization and the chiral order parameter. The internal symmetry
group of these field theories is $SO(3)\times SO(2)$ which is larger than
the $SO(3)$ symmetry of the spin system. The extra $SO(2)$ symmetry
corresponds to rotations in the body fixed frame of the rigid bodies. The
renormalization group analysis of these models\cite{fried,az1,az2} shows 
no novel phases. The system is either in the
Neel ordered phase or in the usual paramagnetic phase with 
exponentially decaying correlation functions and gapped spin one 
magnon excitations. Thus the DTLAF also can be expected not to 
show any novel behaviour near \ki = 1. However, near \ki = 0
we expect some modes other than the Goldstone modes to soften,
reflecting the infinite degeneracy that sets in at \ki = 0. 
The field theory that includes these modes would be appropriate
to study the physics near \ki = 0.

In this paper we motivate a field theory to describe  the Kagome end of 
the model and study its phase structure. We start by finding the 
classical groundstates for different values of $\chi$ in section 2.
We do the spin wave calculation in section 3, systematically 
parametrise the hard and soft fluctuations about the classical ground 
state in the region $0<\chi<1$ and identify the modes that soften when
$\chi \rightarrow 0$. The field theory describing the system near 
$\chi =1$ is derived in section 4. In section 5, we motivate the form of
the field theory near $\chi =0$ that includes large amplitude
fluctuations of the modes that soften in this region.
In  section 6, we integrate out the Goldstone modes and obtain the effective
theory of the new modes. The phases of this effective theory are analysed in
section 7. We summarise our results in section 8. 
 
\section{Classical ground states.}

In this section, we will analyse the ground state of the classical model
.We will show that there are three different types of ground states
corresponding to three ranges of the parameter $\chi$ . The energy of the
classical model can be written as,

\beq E = J \sum_{<i,j> \epsilon K_{B}} \vec {S}_{i}. \vec {S}_{j} + J~
\chi \sum_{<i,j> \not \epsilon K_{B}} \vec {S}_{i}. \vec {S}_{j}
\label{en} \eeq Here, $\vec {S_{i}}$ are vectors satisfying the
constraint $\vec{ S_{i}}.\vec{ S_{i}} = S^{2}$ . $K_{B}$ denotes the set
of bonds that belong to the Kagome lattice. 

We begin with the parameter range $0 < \chi < 2$ . The energy in 
equation (\ref{en}) can be rewritten as,

\beq \frac {E}{J}= \frac {1}{2}\left (1- \frac{\chi }{2} \right) \sum_
{\Delta 
\epsilon K
_{\Delta}}\left (\sum_{i} \vec{S}_{i}\right ) ^2 + \frac{\chi }{4}\sum_
{\Delta \not  \epsilon  K_{\Delta}}\left (\sum_{i} \vec {S}_{i} \right ) ^2
 - \frac
{3S^{2} N}{2}\left (\frac{1+\chi}{2}\right )\label{en2}\eeq
where the sum is over all the triangles that belong to the Kagome
lattice. N is the total number of sites. In the range of $\chi$ 
under consideration, the coefficients of the first two terms in equation
(\ref{en2})
are positive. Thus the energy is minimized by spin configurations 
that satisfy the condition that the net magnetization of
every triangle is zero . It is well known that the unique (upto 
symmetry operations) solution of this constraint is the spiral
state with spiral angle equal to $2\pi /3$ 

Thus, this  so called $ \sqrt{3}\times \sqrt{3}$ state is the  unique, stable
groundstate of the model when $0 < \chi <2 $ . At \ki = 0, of course, 
there are infinitely many other solutions to the constraint 
and the ground state is highly degenerate. 

The ground state energy in this range of \ki  is given by,

\beq E_{G.S} = -\frac {3JNS^{2}}{4} \left (\frac {1+\chi} {2}  \right )
\eeq
Next we look at the range $\chi \geq 2 $ . We rewrite the energy 
as,

 \beq \frac{E}{J} = \sum_{\Delta \not \epsilon K_{\Delta}}\frac{1}{2}\left (
\vec {S}_{1K} +\vec {S}_{2K} + \frac{\chi }{2} \vec{S}_{NK} \right)^{2}
- \frac {3S^{2}N}{2}  \left ( 1+ \frac{\chi ^{2}}{8} \right ) \eeq
Here the sum is over all the triangles that do not belong to  the 
Kagome lattice . $ \vec {S}_{1K} $ and $ \vec {S}_{2K} $   are the spins
at the two Kagome sites and $ \vec {S}_{NK} $ is the spin at the 
non-Kagome site in the centre of every hexagon.

In the range $ 2 \leq \ki \leq 4 $, the quantity ($ \vec {S}_{1K} +
 \vec {S}_{2K} + \frac {\chi} { 2} \vec {S}_{NK} $) can be made to be
equal to zero on every triangle by a non-coplanar spin configuration
described below . Consider any non-Kagome site and let $ \vec {S_{a}} $
and $ \vec {S_{b}} $ be the spins of the $ \sqrt{3}\times \sqrt{3}$ spiral
state on the sites that surround it . Choose,

\beqar \vec{S}_{1K} &=& \cos \theta \vec{S}_{a} + S \sin \theta 
\hat{z} \nonumber \\
\vec{S}_{2K} &=& \cos \theta \vec{S}_{b} + S \sin \theta 
\hat{z} \label{conf1}\eeqar 
If $\theta $ satisfies the equation 

\beq \sin ^{2} \theta = \frac{1}{3}\left ( \frac{\chi ^{2}}{4} -1
\right)\label{sin} \eeq Then we have $ \mid \vec {S}_{1K} +\vec {S}_{2K}
\mid = \frac {\chi}{2} S $ So if we choose

\beqar \vec{S}_{NK} &=& - \frac{2}{\chi } \left ( \vec {S}_{1K} + \vec
{S}_{2K} \right ) \nonumber \\ &=& \frac{2}{\chi } \left ( \cos \theta
\vec {S}_{c} - 2S \sin \theta \hat{z} \right ) \label{conf2} \eeqar Then
the condition $ \vec {S}_{1K} +\vec {S}_{2K} + \frac {\chi} {2}\vec
{S}_{NK} = 0 $ is satisfied in every triangle under consideration .
Equation (\ref{sin}) always has a solution in the parameter range $2 \leq
\chi \leq 4$ . Thus the non-coplanar configuration described in equations
(\ref{conf1}) and (\ref{conf2}) is the stable ground state in the range of
\ki . The ground state energy in this range is given by,

\beq  E_{G.S} = - \frac{3S^2 JN}{2} \left ( 1 + \frac{\chi^{2}}{8}
  \right )\label{en3} \eeq

At \ki = 4, we have $\theta = \pi / 2 $ . All the spins are then 
collinear . The spins on the Kagome lattice point up and the 
others point down . Examining the energy as written in equation (\ref{en3}),
it is clear that this state ($ \theta =\pi /2$) will minimize the 
energy in the range $ \chi \geq 4 $ . The ground state energy in this
range being,

\beq  E_{G.S} = - \frac{3S^2 JN}{2} \left ( \chi -1 \right ) \eeq

In the range $\chi > 2 $, the system has non-zero magnetization.
The average  magnetization per site is given by,

\beqar \vec{M}&=& S \sin \theta \left( \frac{3}{4}-\frac {1}{
\chi } \right ) \hat{z} ~~~~~ 2 \leq \chi \leq 4 \nonumber \\
 &=&    \frac{S}{2} \hat{z}  ~~~~~~~~~~~~~~~~~~~~~~~~~~~~~~ \chi \geq 4 \eeqar 

To summarize, at \ki = 0 the model is exactly the Kagome lattice model
the ground state is infinitely degenerate . As soon as we tune on \ki,
this infinite degeneracy is lifted and we have the $ \sqrt{3}\times
\sqrt{3}$ spiral state as the unique (upto symmetry operations) ground
state . This state remains the ground state until \ki = 2 . The spins then
start lifting off the plane . The spins on the Kagome sites having a $\hat
{z}$ component which is anti parallel to the $\hat {z} $ component of the
spins on the non- Kagome sites . This state is thus a combination of a
spiral and ferrimagnetic state . At \ki = 4 all the spins are collinear
and the transition to the ferrimagnetic state is complete . The
ferrimagnetic state persists for all the values of $ \chi \geq 4$

This completes our analysis of the classical ground states. For the rest 
of the paper we will be focussing our attention on the region 
$ 0 < \chi \leq 2 $ and will be analyzing the fluctuations about the 
 $ \sqrt{3}\times \sqrt{3}$ ground state .
%%%%change made on 8/2/96%%%%%%%%%%%%%%%%%%%%%%%%%%%%%%%%%%%%%%%%%%%%
%\pagebreak
%%%%%%%%%%%%%%%%%%%%%%%%%%%%%%%%%%%%%%%%%%%%%%%%%%%%%%%%%%%%%%%%%%%%%

\section{Spinwave theory}

In this section we do the spin wave analysis of our model hamiltonian in
the region $0 < \chi < 2$. 
The calculation has been done earlier in reference \cite{zeng}. Our aim
here is to compute the gaps as a function of $\chi$ and explicitly 
identify the modes which soften as $\chi \rightarrow 0$.
The hamiltonian is,
\\ \beq H =\sum_{<i,j>}J_{ij}\vec{S_i}.\vec{S_j}\eeq where, $
J_{ij}=\chi$ for i or j belonging to the Kagome lattice and $ J_{ij}=1$
when i and j both lie in the kagome lattice

The unit cell, as shown in fig 1, is a set of 12 points. This is 
commensurate with the periodicity of the DTLAF and the $\sqrt{3} \times
\sqrt{3} $ structure of the classical groundstate.
Adapting our notation to what is suggested by the unit cell structure,  we 
rewrite the hamiltonian, as,

\beq
H=\sum_{Ii\alpha,Jj\beta}\frac{1}{2}J_{Ii\alpha,Jj\beta}Tr[S_{Ii\alpha}S_{Jj\beta}]\eeq
Where we have used the notation, $S_{Ii\alpha} = {1 \over
2}\vec{S}_{Ii\alpha}.\vec{\tau}$,
 $\tau^a$ being the Pauli spin matrices. The index I labels the unit cell.
The set $(i,\alpha)$ label the spins in each unit cell. 
 $\alpha = 0,1,2 $ is the sublattice index. $i=0,..,3$ label the four
different spins of each sublattice in the unit cell. The convention we are
using to label the twelve spins in each unit cell is shown in fig 1. We
then write the spins as, 

%\beq S_{Ii\alpha}={\tilde s}U_{Ii\alpha}^\dagger
\beq S_{Ii\alpha}={\tilde s}\{n_{\alpha} -\frac{i}{\sqrt{\tilde s}}
[w_{Ii\alpha},n_{\alpha}] -\frac{1}{2\tilde
s}[w_{Ii\alpha},[w_{Ii\alpha},n_{\alpha}]]\}
\eeq 
This is the usual Holstein Primakoff transformation and  
$n_{\alpha}={1 \over 2}\vec{n_{\alpha}}
.\vec{\tau},~ \vec{n_{\alpha}}$ being the classical
groundstate spin configuration. 
 ${\tilde s} = \sqrt{S(S+1)}$, so that the magnitudes of the
spins are normalised to $\vec{S}.\vec{S} = S(S+1)$.  
Here, $w_{Ii\alpha}=\frac{1}{2}{\vec w_{Ii\alpha}}.{\vec \tau}$,
 with ${\vec w_{Ii\alpha}}$
being perpendicular to the ground state spin orientation, $\vec n_\alpha$.
Hence $\vec{\wIi} = \tilde P_{Ii\alpha}\hat{\epsilon}_{\alpha}^{1}~+~\tilde 
Q_{Ii\alpha} 
\hat{\epsilon}_{\alpha}^{2} $, with $\hat \epsilon_{\alpha}^1 
,\hat \epsilon_{\alpha}^2,\hat n_{\alpha} $ forming an orthogonal set of
basis vectors for each $\alpha$ and $[\tilde Q_{Ii\alpha},\tilde
P_{Jj\beta}]=i\delta_{IJ}\delta_{ij}\delta_{\alpha \beta}$.
The hamiltonian expanded to the quadratic order in the
fluctuations, \wIi, is given by,
\beq H=\frac{\tilde
s^2}{2}\sum_{r=0,3}J_{Ii\alpha,Jj\beta}^rTr[-\frac{1}{2\tilde s}
\nal\nbet(\wIi^2+\wJj^2)-\frac{1}{ \tilde s}\wIi\nal\wJj\nbet]\eeq we
define the fourier transform as follows,
\beq \wki=\sum_{I}\wIi\exp{(-i\vec{K}.\vec{I})}\eeq
Where I is the unit cell index and $\vec{K}$ takes values in
the  Brilloiun Zone.
Now we can write H in the form,
\beq H=J\tilde s\sum_{K,i\alpha,j\beta}[\tilde P_{Ki\alpha}
 M^{-1}_{i\alpha, j\beta}\tilde P_{-Ki\alpha}+\tilde Q_{Ki\alpha}
K_{i\alpha,j\beta}\tilde Q_{-Ki\alpha}]\eeq
For arbitrary $\chi$ the matrices $M^{-1}$ and K do not commute, hence 
it is not possible to directly diagonalise H. We can define the matrix
$M^{-1}K ~=~ \Omega^2$, and the left and right eigenvectors of this matrix
,$\Psi_L^{n,r}$ and $\Psi_R^{n,r}$,are the normal modes of H and the 
eigenvalues $\omega^2_{n,r}$ are the corresponding energy gaps.
The old variables $\tilde P$ and $\tilde Q$ are written in terms
of the new canonical variables $P $ and $Q$ as follows,
\beqar \tilde P_{K i\alpha}&=&
2\Psi_{R,i \alpha}^{nr}\sqrt{\frac{c'_{n r}}{\omega_{n r}}}
P_{Knr}\\
\tilde Q_{K i\alpha}&=&
2\Psi_{L,i \alpha}^{nr}\sqrt{\frac{c_{n r}}{\omega_{n r}}}
Q_{Knr}\eeqar

The explicit form of M$^{-1}$ and K 
 and our derivation of the normal modes of H are given
in the Appendix-A.
In terms of the new canonical variables $P_{nr}$ and $Q_{nr}$, the Hamiltonian is,
\beq H = \frac{1}{2}\sum_{n r} \omega_{n r}(P_{Kn r}P_{-Kn r}+
Q_{Kn r}Q_{-Kn r}) \eeq 
Explicit  expressions  for the gaps and the left and right eigenvectors
of the matrix $\Omega ^2$ are given in the appendix-B.
The modes (0,0), (0,1), (0,2) are the soft
modes, gapless for all $\chi$, which we shall address as the S-S modes.
The modes (1,0), (1,1), (3,1), (1,2), (3,2), which are hard for
non-zero $\chi$ but become gapless for $\chi=0$, will be referred to as
the H-S modes. The modes (2,0), (3,0), (2,1), (2,2), which remain hard
for all $\chi$ will be referred to as the H-H modes.  Among the H-S
modes, the modes labelled (1,0), (1,1) and (1,2), become
gapless at $\chi=0$  simply because 3 points from each unit cell
decouple from their neighbours  at $\chi=0$. This can be seen by 
looking at the expressions for the corresponding eigenvectors 
at $\chi = 0$, as given in Appendix B.
 Whereas the modes (3,1) and (3,2) are the
ones which truly soften and become gapless at $\chi =0$. 
A look at the contribution from these different modes to the reduction
of the staggered magnetization, gives an idea about how these modes 
affect the physics close to the Kagome end. 

 The staggered magnetization $M_I$ is given by,
\beq M_{I}=\frac{\tilde{s}}{12}\sum_{i,\alpha}U_{Ii\alpha}^\dagger
\frac{\tau^{1}}{2}
U_{Ii\alpha}\eeq where, $ U_{Ii\alpha}= \exp({i \over {\sqrt {\tilde
s}}}w_{Ii\alpha})$ \\ 
Expanding $M_I$ up to terms quadratic in
$w_{Ii\alpha}$ we have, \beq M_{I}=\frac{\tilde
s}{12}\sum_{i\alpha}\frac{1}{2}[ \tau^{1}-\frac{2i}{\sqrt{\tilde s}}
\wIi\tau^{1}-\frac{2}{\tilde s}\wIi^{2}\tau^{1}]\eeq the average
staggered magnetization is,
\beq< M_{I}>=M_{I}^{cl}(1-\bigtriangleup M_I)\eeq where $\bigtriangleup
M_I$ is given by,
\beq \bigtriangleup M_{I} =
\frac{1}{24}\sum_{a,i,\alpha}[<\wIi^{a}\wIi^{a}>-1] \eeq 
The contributions
to $\bigtriangleup M_I$ coming from the hard and soft modes,
plotted as a function of $\chi$  are shown in fig.2. 
The contribution of the
hard modes is seen to dominate close to \ki = 0. This is because
two of the H-S modes start softening in this region and give a large
contribution to $\bigtriangleup$M. This indicates that while deriving the 
low energy effective field theory of this model we should allow for large 
fluctuations of the modes (3,1) and (3,2) near $\chi = 0$.
Hence, near the KLAF end, the theory should be described by 5 parameters 
which include three corresponding to the S-S modes and two corresponding to the 
H-S modes. But before we look at this theory, we take a brief look at 
the physics close to the TLAF.  
\noindent \section{The Field Theory near $\chi = 1$}

As mentioned in the introduction, we expect the lattice spin system to be
well described by a field theory near phase transitions where the physics
is dominated by the low energy, long wavelength modes. This field theory
has previously been derived in reference \cite{domb} for the TLAF. Near
$\chi=1$, the low energy modes are the three Goldstone modes, the S-S
modes. So we must take into consideration the large amplitude fluctuations
of these modes. To do this, we write the spins as, \beq
S_{Ii\alpha}={\tilde s}U_{Ii\alpha}^\dagger n_{\alpha}U_{Ii\alpha}
\label{uspin} \eeq We seperate the hard and the soft modes by rewriting
the $U_{Ii\alpha}$ as, \begin{equation} U_{Ii\alpha} = \expw W_I
\label{param1} \end{equation} Here $ w_{Ii\alpha}$ contains only the H-S
and the H-H modes. In the derivation of the field theory, $w_{Ii\alpha}$
are assumed to be small. There is no assumption about $W_I$ and they can
take any value. The $W_{I}$ correspond to rigid rotations of all the spins
in the unit cell. These are therefore exactly the Goldstone modes.
Therefore, if $w_{Ii\alpha}$ is assumed to be small, we have a
parameterization of the spins such which allows for large fluctuations of
the soft (S-S) modes and small fluctuations of the hard (H-S and H-H)
modes. 

The effective action in the long wavelength, low energy approximation is
obtained by keeping only the terms quadratic in the hard fluctuations, and
then integrating them out. This leaves us with the effective
field theory of the soft modes.  The details of this method of deriving
the field theory will be described elsewhere. The final expression for the
action that we get is, 
\beq 
S=\int d^{3}x \frac{1}{2}\sum_{\mu,a}
\rho_{\mu}^{a}L_{\mu}^aL_{\mu}^{a} 
\label{chi1action} 
\eeq where,
$L_{\mu}^{a}~=~ \frac{1}{2} Tr[\tau^{a}\partial_{\mu}W(x)W(x)^{\dagger}]$,
$ \rho_{0}^{0}~=~\frac{1}{J}\frac{4}{9\sqrt{3}}\frac{(3-\chi)}{\chi(2-\chi)}~,
~ \rho_{0}^{1,2}~=~
\frac{1}{J}\frac{4}{9\sqrt{3}}\frac{(3+7\chi)}{\chi(4+\chi)}~,~ \\
\rho_{i}^{0}~=~JS^2\sqrt{3}~(1+\chi)~$ and
$ \rho_{i}^{1,2}~=~JS^2\sqrt{3}~\frac{\chi(5-\chi)}{(3+\chi)}$, for
$i=1,2$.

Our expressions for the parameters, evaluated at $\chi=1$ coincides with
the values given in reference\cite{domb}. Before ending this section we
describe the symmetries of this model. The original spin hamiltonian is
invariant under the $SO(3)$ spin rotations.  This corresponds to the spins
$S_{Ii \alpha}$ transforming as follows, \beq S^a_{Ii \alpha} \rightarrow
(\Omega_R)^a_b S^b_{Ii \alpha} \nonumber \eeq Where $\Omega_R$ is a
$SO(3)$ matrix. In terms of the matrices $S_{I i \alpha}$, \beq
S_{Ii\alpha} \rightarrow X^\dagger S_{Ii\alpha} X \label{srt} \eeq where X
is the SU(2) representative of the matrix $\Omega_R$. From
equations(\ref{uspin}, \ref{param1}), we see that this corresponds to the
transformation, \beq W(x) \rightarrow W(x)X \label{urt} \eeq $L_{\mu}$ and
hence the action in equation (\ref{chi1action}) are invariant under this
transformation. We refer to this symmetry as the $SO(3)_R$ symmetry. 

In addition the action is also invariant under the transformation, \beq
W(x) \rightarrow Y W(x) \label{ult} \eeq where $Y~ \epsilon~ SO(2)$ and
consists of matrices of the form $ \exp{i \theta \tau^3}$. We refer to
this symmetry as the $SO(2)_L$ symmetry. It acts on the spins as follows,
\beq S^a_{Ii\alpha} \rightarrow (\Omega_L)_{\alpha}^{\beta} S^a_{Ii\beta}
\label{slt} 
\eeq 
where $\Omega_L$ is the $SO(2)$ matrix corresponding to $Y$. 
This transformation is not a symmetry of the lattice model. It only
becomes so in the continuum field theory. The rudiment of this symmetry is
observed in the lattice spin wave hamiltonian as a discrete $Z_3$
symmetry. This comes from the way the translation symmetry of the original
spin hamiltonian is realised and is discussed in more detail in reference
\cite{az1}. The full internal symmetry group of the model is therefore 
$SO(3)_R \times SO(2)_L$. 
   
\noindent
\section{ The field theory near  $\chi=0$ }

We now turn to the low energy physics near the $ \chi = 0 $. 
To include the large amplitude fluctuations of the H-S
modes, we look for a parametrization of the spins as in equation
(\ref{param1}), in which both the S-S and the H-S modes are allowed to have
large fluctuations. We also want the parameterisation to be in terms of quantities
defined over the whole unit cell (i.e. independent of the indices $i$ and $\alpha$)
just as the S-S modes were represented by $W_I$.

The small fluctuations of the classical ground state configuration due to the H-S
modes can be written using equation (\ref{uspin}) as,  
\beq
S_{Ii\alpha}= {\tilde s}n_{\alpha}+i{\sqrt {\tilde s}}[n_{\alpha},w_{Ii\alpha}]
\label{sfluchs}
\eeq
where, $w_{Ii\alpha} = [P^{+} \Psi^{3,1}_{i\alpha} + 
P^{-} \Psi^{3,2}_{i\alpha}]\epsilon^1 _{\alpha}$  
with  $P^+ =(P^-)^{*} = P_1 - iP_2 $. 

We find that if we parameterise the $U_{Ii\alpha}$ matrices in terms of a
unit vector $\hat m_I$ as follows,
\beq 
U_{Ii\alpha} = \exp{i\frac{(i-1)\phi \tau ^3}
{2}} \exp{i \frac{\phi m_I}{2}}
\exp{i \frac{(1-2i)\phi \tau ^3}{4}}
\label{paramhs} 
\eeq
 where, $\phi=2\pi / 3 $, $m_{I}=\hat {m}_{I}.\vec {\tau}$. 
Then the small fluctuations in equation (\ref{sfluchs}) are exactly 
reproduced when $\hat m_I$ is taken to be a small deviation from the 
$z$ axis. Namely, $m_I =  \tau^3 + \pi ^1 \tau ^1
+ \pi ^2 \tau^2 $ and taking upto linear terms in $
\pi^a$. Where $\pi _1 = \frac {1}{3} P_2 $ and $\pi_2 = \frac{1}{3}
 P_1 $.
 Equation (\ref{paramhs}) thus gives a parameterisation, in terms
of a unit vector field, of the large amplitude fluctuations caused by the
H-S modes.

The complete expression for $U_{Ii\alpha}$ including 
the effects of the H-S, the H-H and the S-S modes can be written 
as,
\beq U_{Ii\alpha} = \expwd ~ V_{Ii} ~W_I \label {param2}\eeq
where the $w_{Ii\alpha}$ is expanded in terms of the H-H modes alone
, the $ V_{Ii}$
is given by the R.H.S of equation (\ref{paramhs}) and the 
SU(2) matrix $ W_I$  contains the S-S modes.
The expression (\ref{paramhs}) 
shows that the H-S modes cause a 
deformation of the spin arrangements within the unit cell. 
The fluctuations corresponding to 
$W_I$ cause rigid rotations of the spins within a unit cell 
as before and are the Goldstone modes. 

We now examine the transformation properties of the new fields 
under the symmetries of the theory. 
First we consider the $SO(3)_R$ spin rotation symmetry of the hamiltonian.
The transformation of the spins under this symmetry is given in
equation(\ref{srt}).  This transformation of the spins is obtained if
$W_I$ and $m_I$ transform as follows, \beq W_I \rightarrow W_I X
\label{wrt} \eeq and \beq m_I \rightarrow m_I \label{mrt} \eeq

$\hat{m} $ is therefore a spin singlet.
Next we consider the $SO(2)_L$ symmetry described in section 5. As
mentioned there, this is not a symmetry of the spin system but is however
a symmetry of the low energy, longwavelength field theory near $\chi=1$. We
assume that this symmetry persists near $\chi=0$ also. The transformation of
the spins in equation(\ref{slt}) is obtained if we have, \beq W\rightarrow
Y W \label{wlt} \eeq and \beq m_I\rightarrow Y m_IY^\dagger \label{mlt}
\eeq Equations (\ref{wrt}~-\ref{mlt}) then specify the transformation
properties of the fields under the $SO(3)_R \times SO(2)_L$
symmetry of the low energy theory. 

We now motivate the form of the action that will effectively describe the 
the phases of the DTLAF for small $\chi$. We split up the action as,
\beq
S = S_W[W] +S_{int}[W,m] +S_m[m] 
\label{chi0action}
\eeq
As stated above, we assume that the full symmetry of the model to be $SO(3)_R
\times SO(2)_L$ in the continuum limit. Retaining terms quadratic in the
derivatives, the most general form of the $S_m$ is,
\beq
S_m ~=~\int d^3x \frac{1}{g_2} \partial_\mu m^a 
\partial_\mu m^a+V(m^3) 
\label{seffm}
\eeq
This  action is trivially invariant under the $SO(3)_R$
symmetry since $\hat m$ is a singlet under this symmetry. 
We have taken the derivative terms to be fully $SO(3)_L$ symmetric. We could have
introduced an XY anisotropy but it does not make any qualitative difference in 
the one-loop
approximation we will be working with. $V(m^3)$ however is symmetric only
under $SO(2)_L$. At the classical level, a model defined by $S_m$ has two 
phases.  The disordered $SO(2)_L$ symmetric phase which occurs when $V(m^3)$ is 
minimised at $m^3= \pm 1$ and the ordered $SO(2)_L$ broken phase when it 
is minimised
at $m^3 \ne 1$. For definiteness, we take the potential to be,
\beq
V(m^3)~=~\frac{\lambda_0}{2} (m^3-\eta_0)^2
\label{mpot}
\eeq
Thus for $\eta_0 > 1$, we have the symmetric phase (classically), there are two
modes with equal gaps which are equal to $g_2\lambda_0(\eta_0-1)/2$. For $\eta_0 < 1$,
the $SO(2)_L$ symmetry is broken. There is one gapless Goldstone mode and the
other mode has a gap equal to $g_2\lambda \sqrt {(1-\eta^2_0)}/2$. The spin wave
analysis in section 3 showed that the two H-S modes had equal gaps which went
to zero as $\chi \rightarrow 0$. We therefore take the unrenormalised value of
$\eta_0$ to be equal to 1.

The general form of $S_W$ that retains terms quadratic in the derivatives and 
consistent with the symmetries of the theory is given by equation (\ref {chi1action}).
To motivate the form of the interaction term, we note that the deviation of $m^3$ from
$\pm 1$ implies that the spin configuration is nonplanar. To see this, we define vectors
$\hat C_{Ii}$ as,
\beq
C_{Ii}= -{2i \over 3 \sqrt 3} \sum_\alpha [S_{Ii\alpha},S_{Ii\alpha+1}]
\label{chivec}
\eeq 
where as usual $C_{Ii}=\hat C_{Ii}.\vec \tau$. $\hat C_{Ii}$ is the normal to the
plane on which the  3 spins labelled by a particular value of $i$ lie. Using equation
(\ref{paramhs}) we have
\beq
C_{Ii}= e^{i \frac{(i-1)\phi \tau ^3} {2}} e^{i \frac{\phi m_I}{2}}
~\tau^3~ e^{-i \frac{\phi m_I}{2}} e^{-i \frac{(i-1)\phi \tau ^3}
{2}}
\label{chiform}
\eeq
It is clear that when $\hat m$ deviates from $\hat z$, the vectors $\hat C_{Ii}$ are non
coplanar. It is known that
nonplanarity of the background spin configuration makes the gapless spin waves stiffer
\cite{shend}. We therfore write down an interaction term of the form,
\beq
S_{int}= \int d^3x f(m^3) L_{\mu}^a L_{\mu}^a 
\label{sint}
\eeq
Where $f(m^3)$ increases as $|m^3|$ decreases. For simplicity,
 we take $f(m^3)~=~-\alpha
(m^3)^2$ with $\alpha > 0$.

%\frac{1}{g_2}\partial_\mu m^a \partial_\mu m^a ~-~h~m^3~ +~\frac{\lambda}{2}
%~ (m^3)^2\label{chi0action}\eeq
%Where the $L^a_\mu$ and $m^a$ are defined as before and $\frac{1}{\tilde{g_1}}=
%\frac{1}{g_1} - \alpha(m^3)^2$. 
%\beqar S_W &=& \int d^3x \frac{1}{g_1} L_{\mu}^a L_{\mu}^a
%\\

\section{Integrating out the W fields}

We now  investigate the phases of the field theory that has been proposed in
the previous section. In particular we are interested in seeing if there is a phase 
in which the $SO(2)_L$ symmetry is broken and the $SO(3)_R$ spin symmetry is unbroken. At 
values of $\chi$ where the system is effectively described by a field theory of form given
in equation (\ref{chi1action}), it is known that this does not happen \cite{fried,az1,az2}.  
However, as mentioned in the previous section, this does occur in the field theory given in
equation (\ref{chi0action}) at the classical level if $\eta_0 < 1$. We have also argued
that the unrenormalised value of $\eta_0$ is equal to 1. The potential $V(m^3)$ in equation
(\ref{mpot}) will get modified by the fluctuations of both the $W$ and the $\hat m$ fields. 
In this section we will integrate out the $W$ fields and compute the above mentioned change. 
We then investigate the effect of the $\hat m$ fluctuations by a renormalization group 
analysis of $S_m$ in the next section. 

If $\Delta V(m^3)$ is the change in the bare potential due to the $W$ fluctuations, then we
have,
\beq
e^{-\int_x \Delta V(m^3)}~=~\int_W e^{-(S_W[W]+S_{int}[W,m^3])}
\label{delpotdef}
\eeq
$S_W$, as stated earlier, is of the form given in equation (\ref{chi1action}). It is known
\cite{az1}, that the two renormalised spin wave velocities tend to become equal. So we make
the simplifying assumption of space-time isotropy and work with $S_W$ of the form,
\beq 
S_W~=~ \int d^3x \frac{1}{g_1} \sum_{a=1}^2L^a_\mu L^a_\mu ~+~
\frac{1}{g_3} L^3_\mu L^3_\mu 
\label{waction}
\eeq

We first consider the weak coupling regime when $g_1,g_3~<<~1$. In this regime the $W$
fields are ordered and the $SO(3)_R$ symmetry is broken. The $W$ integration can be done
semiclassically and we get,
\beq
\Delta V(m^3) = -(g_1+g_2)\alpha (m^3)^2
\label{delpotw}
\eeq
Thus in the weak coupling regime, where the $W$ field is ordered, we have $\eta_0 \to
\eta_0 / (1-{(g_1+g_2)\alpha) \over \lambda})$. Therefore, in the ordered phase, the $W$ 
field fluctuations increase the value of $\eta_0$. 

Next we consider the strong coupling regime, $g_1,g_3 >>1$, where the $W$ fields are
disordered. In this regime, the $SO(3)_R$ symmetry is unbroken. We first rewrite the theory
in terms of a set of three orthogonal vectors defined as below
\beq
\phi_r^a~=~{1 \over 2\gamma_a}tr (\tau_a W^{\dagger} \tau_r W)
\label{phidef}
\eeq
Where $\tau_a$ are the Pauli matrices, the indices $a,r~=~1,2,3$ and 
${1 \over \gamma_1}={1 \over \gamma_2}~=~{1 \over g_1}-\alpha(m^3)^2,~{1 \over \gamma_3}
~=~{1 \over g_3}-\alpha(m^3)^2$. From the
definition, $\phi_r^a$ satisfy the orthogonality conditions,
\beq
\sum_{r=1,3}\phi_r^a \phi_r^b ~=~{1 \over \gamma_a} \delta^{ab}
\label{orthrel}
\eeq
The action in equation(\ref{delpotdef}) can be rewritten in terms of these fields as,
\beq
S_W+S_{int}~=~\int d^3x \sum_{a=1,3}\sum_{r=1,3}
[ \partial_{\mu} \phi_r^a \partial_{\mu} \phi_r^a
+ i \Lambda^{ab} (\phi_r^a \phi_r^b -{1 \over \gamma_a}\delta^{ab})]
\label{phiaction}
\eeq
$\Lambda^{ab}$ are Lagrange multiplier fields that impose the constraint in equation
(\ref{orthrel}).
The action is quadratic in the $\phi$ fields and they can be integrated out. We are then
left with the integral over the $\Lambda$ fields with an effective action given by,
\beq
S_{eff}= \frac{1}{2} Tr ln (-(\partial_\mu^2)\delta^{ab} + i\Lambda^{ab})
-i\int d^3x {1 \over \gamma_a}\Lambda^{aa}
\label{lamaction}
\eeq
We now do the integration over the $\Lambda$ fields in the saddle point approximation. This
is a well known technique that is exact in the large N limit where the index $r$ runs from 1
to N and the coupling constants are suitably rescaled. The saddle point equations are,
\beq
\int {d^3k \over (2\pi)^3}({1 \over k^2+i\Lambda})^{ab}= {2 \over \gamma_a}\delta^{ab}
\label{speq}
\eeq
In the strong coupling regime, the solution is $i\Lambda^{ab}=M^2_a\delta^{ab}$, where
$M^2_a$ are non-zero. The $W$ fields are thus disordered with  correlation lengths 
$\xi_a=M^{-1}_a$. In the saddle point approximation then, $\Delta V(m^3)$ is given by,
\beq
\Delta V(m^3)~=~ \frac{1}{2} Tr ln (-(\partial_\mu^2)+M^2_a)\delta^{ab})
-\int d^3x {1 \over \gamma_a}M^2_a
\label{delpots}
\eeq
Here $M_a$ are the solutions of the saddle point equations. Thus both $\gamma_a$ and $M_a$
in equation (\ref{delpots}) are functions of $m^3$. To see the form of the dependence of
$\Delta V(m^3)$ on $m^3$ in equation (\ref{delpots}), we differntiate it with respect to
$(m^3)^2$. Using the saddle point equation (\ref{speq}), we obtain,
\beq
{\partial \Delta V \over \partial (m^3)^2}~=~\sum_a M^2_a \alpha
\label{delpotdiff}
\eeq
Thus $\Delta V$ is a monotonically increasing function of $(m^3)^2$ and is minimised at
$m^3=0$. Therefore, in the strong coupling regime, the $W$ fluctuations decrease the value
of $\eta_0$.

The important conclusion that we draw from the above results is that in the weak coupling
regime, where the $W$ fields are ordered and the $SO(3)_R$ symmetry is broken, the $W$
field fluctuations increase $\eta_0$ and therefore tend to restore the $SO(2)_L$ symmetry.
On the other hand in the strong coupling regime when the $W$ fields are disordered, and the
$SO(3)_R$ symmetry is unbroken, the fluctuations decrease the value of $\eta_0$ and hence
tend to break the $SO(2)_L$ symmetry.

%herex

\section{The $\hat m$ field fluctuations}

In this section, we investigate the effects of the $\hat m$ field fluctuations by a
renormalization group analysis of $S_m$.  The theory has three coupling constants, 
$g_2, \lambda$ and $\eta$. The one loop
renormalization group equations that govern their flow can be computed using standard
techniques. They turn out to be,  
\beqar
\frac{\partial g_2}{\partial l}~&=&~ -g_2 +g_2^2 \\
\frac{\partial\lambda}{\partial l}~&=&~ 3\lambda (1-g_2)\\
\frac{\partial \eta}{\partial l}~&=&~ 2g_2(1+\eta)
\label{rgeq}
\eeqar
These equations can be explicitly solved to get,
\beqar
g_2&=&\frac{g_{20}\exp(-l)}{1-g_{20}(1-\exp(-l))}\\
\lambda&=&\lambda_0~(1-g_{20}(1-\exp(-l)))^3~\exp(3l)\\
1+\eta&=&(1+\eta_0)(1-g_{20}(1-\exp(-l)))^{-2} 
\label{rgsol}
\eeqar
When $g_{20} < 0$, $g_2$ flows to 0 and $\lambda$ flows to $\infty$.
Therefore, in this range of $g_{20}$, The $SO(2)_L$ symmetry will be broken
if $\eta(\infty) < 1$ and will be intact otherwise. The phase boundary is
then given by the equation,
\beq
(1+\eta_0)~=~2(1+g_{20})^2
\label{phbndry}
\eeq
Thus the $\hat m$ field fluctuations do not succeed in restoring the 
$SO(2)_L$ symmetry everywhere. There is a region of the couplings 
$g_{20}$ and $\eta_0$ shown in figure (3) for which the $SO(2)_L$ 
symmetry remains broken. 

We can use the vectors $\hat C_{Ii}$ defined in equation(\ref{chivec})
to define an order parameter for this transition in terms of the spins.
We define
\beq
\Psi=1-{\hat C}_{Ii}.{\hat C}_{Ii+1}
\label{lopdef}
\eeq
$\Psi_I$ can be expressed in terms of $W_I$ and $\hat m_I$. It is independent of 
$W_I$ (since it is a spin singlet) and is equal to
\beq
\Psi_I={9 \over 8}sin^2(\theta)(3cos^2(\theta)+1)
\label{lopval}
\eeq
So $\Psi$ is 0 in the $SO(2)_L$ unbroken phase and is $\ne 0$ in the broken phase. 

\section{Summary}

To summarize, we have studied the DTLAF which interpolates between the
TLAF and the KLAF. The classical ground state, in the region, $0<\chi\le
2$, is the $\sqrt 3 \times \sqrt 3$ Neel ordered state. 

We have computed the spin wave spectrum of the DTLAF in the above
mentioned regime. There are 3 gapless Goldstone modes which we have
called the S-S modes. 5 have gaps which go to zero as $\chi \rightarrow
0$, the H-S modes. The remaining 4 have a gap throughout the region and we
have called them the H-H modes. The S-S and the H-S modes are important for 
the field theory that would describe the low energy long wavelength physics 
of the system in the small $\chi$ region. There are 3+5=8 such modes. In the
$\chi \rightarrow 0$ limit, the system decouples into the KLAF and a bunch
of decoupled individual spins (3 per unit cell) sitting of the triangular
lattice sites that do not belong to the Kagome lattice. If we are
interested only in the spins of the KLAF, then we have only 5 of these 8
modes are left. We then allowed for large fluctuations of these modes and
found that they can be thought of as fluctuations of an order parameter
that takes values in $SO(3) \times S_2$. Namely a $SO(3)$ matrix
 W  and a
unit vector $\hat m$. 

Based on this, we have written down an effective action in terms of 
the fluctuations of W and $\hat m$. We assume that the symmetry of the
theory is enhanced  to $SO(3)_R \times SO(2)_L$ in the continuum limit 
as it happens in the $\chi=1$ end. We also allow for a simple 
interaction between these fields that is consistent with symmetry
requirements and other known facts about the system. We have then 
integrated out the $W$ fields in the weak and strong coupling regimes,
and have analysed the resulting effective theory of the $\hat m$ fields 
by a one loop renormalization group calculation. 

We find that in the region where $g_1$ is small and the $W$ field is
ordered, the $SO(2)_L$ symmetry remains unbroken and the gap of the $\hat
m$ field is increased due to quantum fluctuations.
 In the regime $g_1 >1$ 
where the spins are quantum disordered and the SO(3)$_R$ spin
symmetry is unbroken, the  $W$ field fluctuations drive
the $\hat m$ system to a phase where the $SO(2)_L$ symmetry between 
is broken and there exists one gapless Goldstone mode in the spectrum.

This is our proposal for the mechanism that produces a gapless excitation
while keeping the symmetries of the hamiltonian intact. While we have shown
the existence of this phase in the continuum field theory, we cannot say if
the spin system is actually realised in this phase. To answer this question
within the framework we are working in, we have to derive the values
of the coupling constants in the field theory from the spin system as we
have done in the $\chi=1$ end. This work is in progress.
\vskip 2cm
\centerline{\bf APPENDICES} \vskip 1cm \appendix \section {The Matrices
$M^{-1}$ and $K$} \beqar
M^{-1}_{i\alpha,j\beta}&=&\frac{1}{2}[A_{i,j}\otimes I_{\alpha,\beta} +
2(B^{0}+B^{1}+B^{2} )_{i\alpha,j\beta}]\nonumber \\
K_{i\alpha,j\beta}&=&\frac{1}{2}[A_{i,j}\otimes I_{\alpha,\beta} -
(B^{0}+B^{1}+B^{2})_ {i\alpha,j\beta}]\nonumber \eeqar
 %changes made on 8/2/96%%%%%%%%%%%%%%%%%%%%%%%%%%%%%%%%%%%%%%%%%%%%%%%
% \\[2.5 mm] 
where,\\
%\\[2.5mm]
 $\begin{array}{l} ~~~~~~~~~~~~~~~~~~A_{i,j}=\left [
\begin{array}{cccc}
         \chi +2 & 0 & 0 & 0\\
          0 & \chi +2 & 0 & 0 \\
          0 & 0 & \chi +2 & 0 \\
          0 & 0 & 0 & 3\chi \\
           \end{array}\right ]\\
\\[2.5 mm]
\mbox{the matrices $ B_{0,1,2} $ are given by, }\nonumber\\
\\[2.5mm]
~~~~~~~~~~~~~~~~~~B_{0}= \frac{1}{2} \left[ \begin{array}{ccc}
			0 & \tilde{B} & \tilde{B}^{T} \\
			\tilde{B}^{T} & 0 & \tilde{B}\\
			\tilde{B} & \tilde{B}^{T} & 0\\
                        \end{array} \right]\\
\\[2.5mm]
\mbox{and}\\
\\[2.5mm]
~~~~~~~~~~~~B_{1} +B_{2}= \frac{1}{2} \left [ \begin{array}{ccc}
		 0 & \tilde{B_{1}} & \tilde{B_{2}}^{\dagger} \\
		 \tilde{B_{1}}^{\dagger} & 0 & \tilde{B_{3}} \\ 
	         \tilde{B_{2}}& \tilde{B_{3}}^{\dagger}& 0 
		             \end{array} \right ]  
\end{array}$
\\[2.5mm]
\newpage
$\begin{array}{l}
\\[2.5mm]
\mbox{where,}
\\[2.5mm]
~~~~~~~~~~~~~~~\tilde{B_{i,j}}=\left [ \begin{array}{cccc}
		1&1&0&\chi \\
		0&1&1&\chi \\
		1&0&1&\chi \\
		\chi & \chi & \chi & 0\\
		\end{array} \right ]\\
\end{array}$
\\[2.5mm]
$\begin{array}{l}
\\[2.5mm]
\mbox{and}
\\[2.5mm]
~~~~~~~\tilde{B_{1}}(K_{1},K_{2},K_{3})= \left [ \begin{array}{cccc}
		   0 & 0 & 0 & \\
		   0 & \exp{(-iK_{2})} - 1 & 0 & 0 \\
	  	   0 & 0 & \exp{(iK_{2})} -1 & \chi (\exp{(-iK_{3})} -1) \\
		   0 & \chi (\exp{(iK_{1})} -1) & 0 & 0 \\
		   \end{array} \right ]\\
\\[2.5mm]
~~~~~~~\tilde{B_{2}}(K_{1},K_{2},K_{3})= \tilde{B_{1}}(K_{2},K_{3},K_{1})\\
\\[2.5mm]
~~~~~~~\tilde{B_{3}}(K_{1},K_{2},K_{3})= \tilde{B_{1}}(K_{3},K_{1},K_{2})
\end{array}$
\\[5mm]		   

%%%%%%%%%%%%%%%%%%%%%%%%%%%%%%%%%%%%%%%%%%%%%%%%%%%%%%%%%%%%%%%%%%%%%%
%
%  Begin shift to Appendix
%
The matrices $M^{-1}$ and K do not commute for arbitrary $\chi$, so in order
to diagonalise the hamiltonian we define the normal modes as the  left and
right eigenvectors of the matrix $\Omega^2 =KM^{-1}$ as follows,
 
\beqar \Omega^2\Psi_{R}^{n r}&=&\omega^2_{n r} \Psi_{R}^{n r} \\
 \Omega^{2^T}\Psi_{L}^{n r}&=&\omega^{2}_{n r} \Psi_{L}^{n r} \eeqar
where the indices (n,r) are similiar to the indices$ (i,\alpha)$
The $\Psi_{L,R}$ have the properties,
\beqar(\Psi_{L}^{n r},\Psi_{R}^{n' r'})&=&\delta_{n,n'}\delta_{r,r'} \\
\sum_{n,r}\Psi_{L}^{n r}\Psi_{R}^{n r^T} &=& 1 \eeqar
Also,
\beqar M^{-1}\Psi_{R}^{n r}=c_{n r}\Psi_{L}^{n r} \\
 K\Psi_{L}^{n r}=c'_{n r}\Psi_{R}^{n r} \\
\omega^2_{n r}=c_{n r}c'_{n r} \eeqar
%
% End shift to Appendix
% begin second shift to appendix
The \wki can be expressed in terms of these normal modes as follows,
\beqar \tilde P_{Ki\alpha}&=&\sum_{n r}2
\sqrt{\frac{c'_{n r}}{\omega_{n r}}} \Psi_{R i \alpha}^{n r}P_{n r} \\
       \tilde Q_{Ki\alpha}&=&\sum_{n r}2
\sqrt{\frac{c_{n r}}{\omega_{n r}}} 
\Psi_{L i \alpha}^{n r}Q_{n r} \eeqar
This is a canonical transformation since
\beq [Q_{n r},P_{n' r'}]=i\delta_{n,n'}\delta_{r,r'} \eeq 
Written in terms of $P_{n,r}$ and $Q_{n,r}$ the Hamiltonian is,
%\beq H=\sum_{n,r}\frac{1}{8}[c_{n r}P_{n r}^{2} + c'_{n r}Q_{n r}^{2}]
%\eeq We can rescale the $P_{n,r}$ and $Q_{n,r}$ to write this in a
%standard form, as,

\beq H = \frac{1}{2}\sum_{n r} \omega_{n r}(P_{Kn r}P_{Kn r}+
Q_{Knr}Q_{-Kn r}) \eeq 
 
%\beqar \bar{P_{n r}}^2 &=&\frac{1}{4}\frac{\omega_{n r}}{c'_{n r}}P_{n
%r}^2 \\ \bar{Q_{n r}}^2 &=&\frac{1}{4}\frac{\omega_{n r}}{c_{n r}}Q_{n
%r}^2 \eeqar 
%end second shift into appendix
%%%%%%%%%%%%%%%%%%%%%%%%%%%%%%%%%%%%%%%%%%%%%%%%%%%%%%%%%%%%%%%%%%%%%%%

\section{The Eigenvalues and Eigenvectors of $\bigtriangleup \Omega^{2}$}
As we saw in Appendix A, the matrices M$^{-1}$ and K occur naturally as  
direct products of certain $4\times 4$ and $3\times 3$ matrices and so the eigen
vectors also have this form and this is written as follows.

\noindent $\Psi_{L,R i,\alpha}^{n r} = \Phi_{L,R,i}^{n r} 
X_{\alpha}^{r} $ where,\\ 

$
\begin{array}{l@{\hspace{1cm}}l@{\hspace{1cm}}l}
X^{0} = \frac{1}{\sqrt {3}} \left ( \begin{array}{c} 1\\1\\1
 \end{array} \right ) &
X^{1} = \frac{1}{\sqrt {3}} \left ( \begin{array}{c} 1\\\alpha\\
\alpha ^{2} \end{array} \right ) &
X^{2} = \frac{1}{\sqrt {3}} \left ( \begin{array}{c} 1\\\alpha ^{2}\\
\alpha \end{array} \right ) 
\end{array}
$\\
where,
$ \alpha ^{3}= 1 $ \\
The $ \Phi_{L,R,i}^{n r} $ are given by,\\
$
\begin{array}{l@{\hspace{3cm}}l} 

\\[2.5mm]
\Phi_{R}^{0 0 } = \frac {1}{3 - \chi }\left ( \begin{array} {c} \chi \\
\chi \\ \chi \\ 6 - 5 \chi  \end{array} \right )& 
\Phi_{L}^{0 0} = \frac {1}{2 }\left ( \begin{array} {c} 1 \\ 1 \\ 1 \\
1  \end{array} \right )\\
\\[2.5mm]

\Phi_{R}^{1 0 } = \frac {1}{2 \sqrt{3} }\left ( \begin{array} {c} -1 \\
-1 \\ -1 \\ 3 \end{array} \right )&                      
\Phi_{L}^{1 0} = \frac {1}{\sqrt{3} (3 - \chi) }\left ( \begin{array}{c}
5 \chi -6 \\ 5 \chi - 6 \\ 5 \chi - 6 \\ 3 \chi \end{array} \right )\\
\\[2.5mm]

\Phi_{R}^{2 0 } = \frac {1}{\sqrt{3} }\left ( \begin{array} {c} 1 \\ 
\alpha \\ \alpha ^{2} \\ 0  \end{array} \right )&                      
\Phi_{L}^{2 0} = \frac {1}{\sqrt{3} }\left ( \begin{array} {c} 1 \\ 
\alpha \\ \alpha ^{2} \\ 0  \end{array} \right )\\
\\[2.5mm]

\Phi_{R}^{3 0  } = \Phi_{R}^{2 0 *}&                      
\Phi_{L}^{3 0 } = \Phi_{L}^{2 0 *}

\end{array}
\\
\begin{array}{l@{\hspace{3cm}}l}
\\[2.5mm]
\Phi_{R}^{0 1 } = \frac {1}{2 }\left ( \begin{array} {c} 1 \\
1 \\ 1  \\ 1  \end{array} \right )&                   
\Phi_{L}^{0 1} = \frac {1}{(7 \chi + 3) }\left ( \begin{array} {c} 5\chi 
\\ 5\chi \\  5\chi  \\ (6- \chi)  \end{array} \right )\\
\\[2.5mm]  

\Phi_{R}^{1 1 } = \frac {1}{\sqrt{3}(3 +7\chi)}\left ( \begin{array}
 {c} \chi - 6 \\ \chi - 6 \\ \chi - 6 \\ 15 \chi \end{array} \right )&
\Phi_{L}^{1 1} = \frac {1}{2 \sqrt{3}}\left ( \begin{array} {c} -1 \\ -1
\\ -1 \\ 3 \end{array} \right )\\ \\[2.5mm]

\Phi_{R}^{2 1 } = \frac {1}{\sqrt{3} }\left ( \begin{array} {c} 1 \\ 
\alpha ^{2} \\ \alpha \\0  \end{array} \right )&                    
\Phi_{L}^{2 1} = \frac {1}{\sqrt{3} }\left ( \begin{array} {c} 1 \\ 
\alpha ^{2} \\ \alpha \\0  \end{array} \right )\\ 
\\[2.5mm] 

\Phi_{R}^{3 1  } = \Phi_{R}^{2 1 *}&                     
\Phi_{L}^{3 1 } = \Phi_{L}^{2 1 *}
 
\\[2.5mm]  
%changes made on 8/2/96      %%%%%%%%%%%%%%%%%%%%%%%%%%%%%%%%%%%%%%%%

\Phi_{R}^{i 2  } = \Phi_{R}^{i 1 *}&
\Phi_{L}^{i 2 } = \Phi_{L}^{i 1 *}
 
\\[3mm] 
%%%%%%%%%%%%%%%%%%%%%%%%%%%%%%%%%%%%%%%%%%%%%%%%%%%%%%%%%%%%%%%%%%%%%

\end {array}\\
\noindent \mbox{and}\\ 
\begin{array}{lll}
\\[3mm]

c_{0 0}= \frac{9 \chi (2 - \chi)}{3 - \chi}&   
c'_{0 0}=0& 
\omega^{2}_{0 0}=0\\
\\[3mm]

c_{1 0}=\frac{3 - \chi}{4}  &    
c'_{1 0}=2 \chi &  
\omega^{2}_{1 0}=\frac{\chi (3 - \chi)}{2} \\
\\[3mm]

c_{2 0}= \frac{3 + \chi}{2} &   
c'_{2 0}= \frac{3 + 2 \chi}{4}&  
\omega^{2}_{2 0}=\frac{(3+\chi )( 2\chi + 3)}{8} \\
\\[3mm]

c_{3 0 }=c_{2 0} &   
c'_{3 0 }=c'_{2 0}&  
\omega^{2}_{3 0 }= \omega^{2}_{2 0} 
\\[3mm]

%changes made on 8/2/96       %%%%%%%%%%%%%%%%%%%%%%%%%%%%%%%%%%%%%%%
\end{array}
\\
\begin{array}{lll}
\\[5mm]

c_{0 1 }= 0 &   
c'_{0 1 }=\frac{9\chi (4+\chi)}{2(7\chi + 3)}&  
\omega^{2}_{0 1 }=0 \\
\\[5mm]

c_{1 1 }= 2\chi &   
c'_{1 1 }=\frac{(3 +7\chi )}{8}&  
\omega^{2}_{1 1 }=\frac{\chi ( 3 + 7\chi)}{4} \\
\\[5mm]

c_{2 1 }=\frac{\chi +3}{2} &   
c'_{2 1 }=\frac{\chi + \frac{3}{2}}{2}&  
\omega^{2}_{2 1 }=\frac{(\chi +3) (2\chi +3)}{ 8} \\
\\[5mm]

c_{3 1 }=\frac{\chi}{2}  &   
c'_{3 1 }=\frac{(\chi +3)}{2}&  
\omega^{2}_{3 1 }=\frac{\chi(\chi +3)}{4}  \\
\\[5mm]

c_{i 2 }=c_{i 1} &   
c'_{i 2 }=c'_{i 1}&  
\omega^{2}_{i 2 }= \omega^{2}_{i 1}~~~~~~~~~~~~~~~i=0,...3
\\[5mm] 
\end {array} $ \\ 

\newpage
 %LIST OF REFERENCES 

%%%%%%%%%%%%%%%%%%%%%%%%%%%%%%%%%%%%%%%%%%%%%%%%%%%%%%%%%%%%%%%%%%%%
\newpage
%%%%%%%%%%%%%%%%%%%%%%%%%%%%%%%%%%%%%%%%%%%%%%%%%%%%%%%%%%%%%%%%%%%%
\centerline{\bf figure captions}
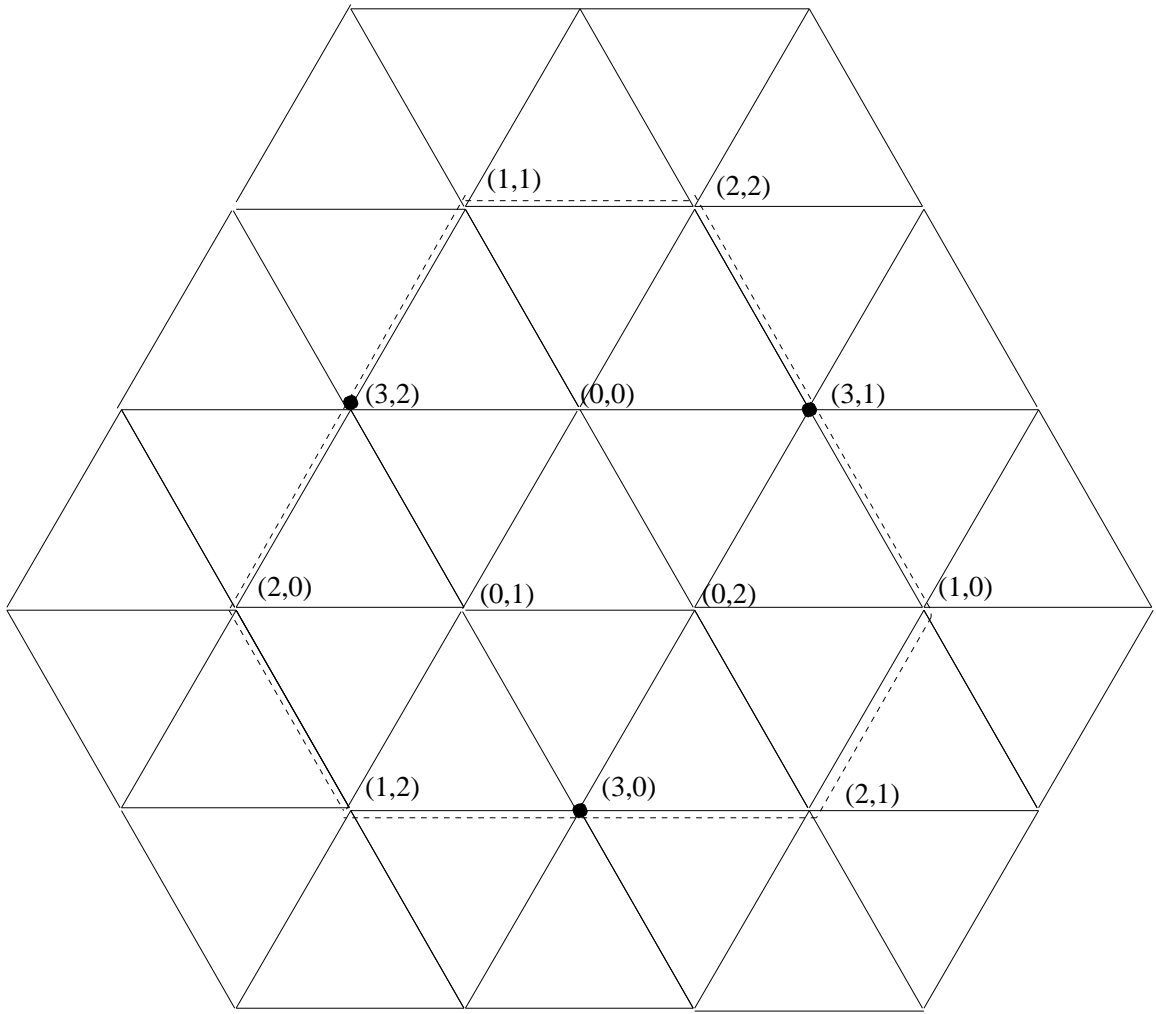
\noindent fig.1: A section of the triangular lattice showing the unit
cell
consisting of 12 points and the
labelling of the points within a unit cell. The non-Kagome points
labelled (3,$\alpha$) are marked with dark spots
and the bonds connected to them are the non-kagome bonds.\\
\\
\\
fig.2: Reduction in the staggered magnetization, $\Delta$M, as a 
function of $\chi$ 
and the contributions to this from the S-S modes ($\Delta$ SM) and
from
the H-H and H-S modes ($\Delta$ HM). Near $\chi =0$ 
  $\Delta$ HM is seen to dominate over $\Delta$ SM.\\
\\
\\
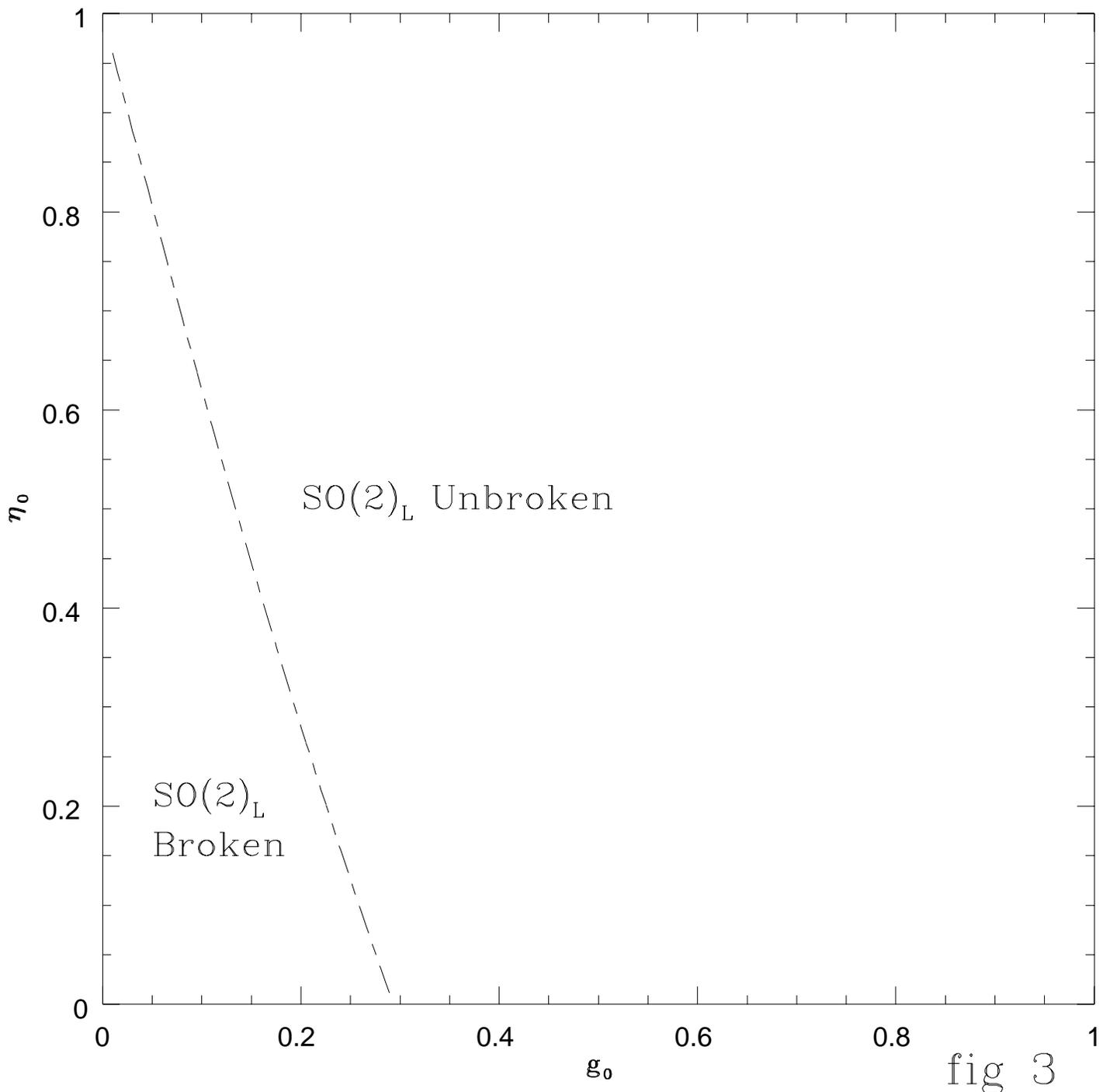
fig.3: Phase diagram showing the boundary   dividing 
the $SO(2)_L$ broken phase from the unbroken phase.

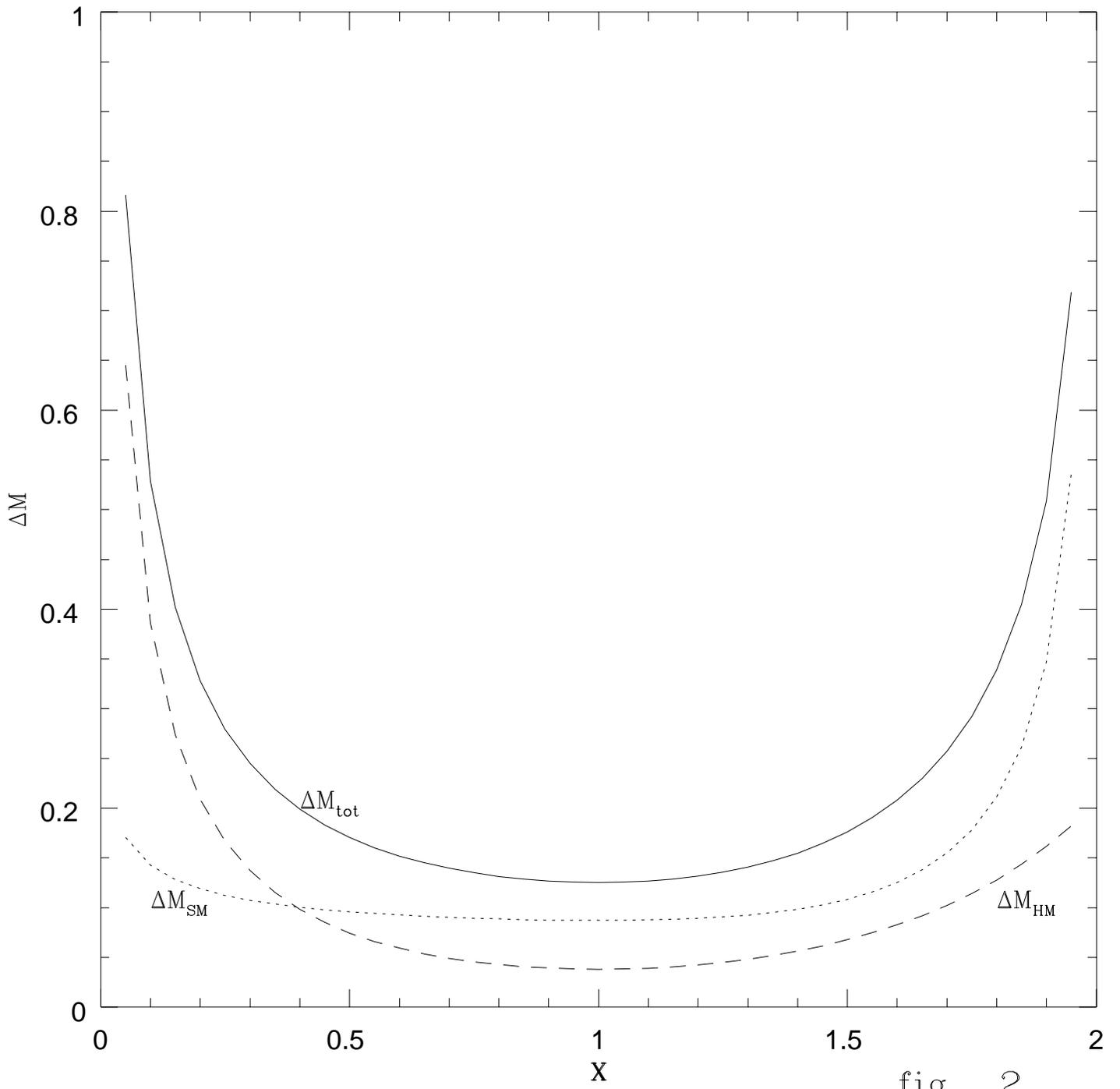
\end{document}